\documentclass[preprint,floatfix,onecolumn,superscriptaddress]{revtex4-1}
\usepackage[dvips]{graphicx}
\usepackage[english]{babel}
\usepackage{amsmath}
\usepackage{amssymb}
\usepackage{tensor}

\newcommand{\vk}{\mathbf{k}}
\newcommand{\vn}{\mathbf{n}}

\newcommand{\be}{\begin{eqnarray}}
\newcommand{\ee}{\end{eqnarray}}
\newcommand{\p}{\partial}

\newcommand{\dc}{c^{\dagger}}

\def\ep#1{\langle #1 \rangle}

\begin{document}

\title{Exact Solutions of Topological Superconductor Model with Hubbard Interactions}

\author{Yiting Deng}
\affiliation{College of Physics, Sichuan University, Chengdu, Sichuan 610064, China}

\author{Yan He}
\affiliation{College of Physics, Sichuan University, Chengdu, Sichuan 610064, China}
\email{heyan$_$ctp@scu.edu.cn}

\begin{abstract}
We study a two-dimensional model of topological superconductor with equal spin pairing and repulsive Hubbard interaction. When the pairing gap equals to the hopping constant, half of the spectrum of this model are flat bands, which makes this model exact solvable. The band structure and topological properties of the exact solutions of the interacting model are analyzed in details. It is found that the ground state corresponds to a staggered distribution of the conserved quantities.
\end{abstract}

\maketitle

\section{Introduction}

The strong correlated systems have been a difficult subject in condensed matter physics for decades \cite{Fradkin-book}. Due to the lack of small parameters, the traditional methods such as perturbation expansion or mean field approximation cannot provide reliable calculations for systems with strong enough couplings. In these situations, exact solvable models are very valuable for understanding the behaviors in strong coupling limit. Unfortunately, most exact solvable models are limited to spatial dimension one and they are usually solved by the Bethe ansatz which is quite complicated mathematically \cite{Bethe,Andrei,karabach1997introduction,Betheansatz}. Exact solvable models in higher dimensions are certainly worth close studies.

The past decade witnessed the rising interests in the study of topological matters \cite{Kane-review,Zhang-review}. This new trend also inspired new ideas on constructing exact solvable models in dimension two (2D) \cite{azaria1987coexistence,sengupta2008exact}. A famous example is the Kitaev model \cite{Kitaev-model,baskaran2007exact,mandal2020introduction,miao2017exact} on the honeycomb lattice which supports nontrivial topological order. Although its original form is a spin model, if one rewrite the Kitaev model in terms of Majorana fermions, then it is evident that there are infinite many conserved quantities, which makes this model exact solvable. More recently, this line of developments has been revived and a large class of exact solvable BCS-Hubbard models have been proposed in a series of works \cite{TKNg2018:prl,TKNg2019:prb,Miao2019:prb,Miao2017:prl}. These types of models are studied under the name of the Falicov-Kimball model many years ago. The common feature of these modes is the appearance of flat bands for certain appropriately chosen parameters. There exist many quadratic terms of flat band fermions which can serve as the infinite conserved quantities. This will transfer the four-fermion interacting Hubbard term to a quadratic fermion term. Then the interacting model can be easily solved as a simple quadratic fermion model. Other than the exact solvability, these models also provide a platform to study the interplay between the strong correlations and topology \cite{Ezawa2018:prb,Ezawa2017:prb}.

In the present paper, we propose a topological superconductor model with equal spin pairing, based on the Qi-Wu-Zhang model \cite{QWZ,asboth2016short} which is a Chern insulator. In section \ref{sec-construct}, we present the a detailed construction of the model Hamiltonian. We also point out that, for certain choice of pairing gap, one half of its energy bands become flat bands. In section \ref{sec-Hubbard}, we transfer the model Hamiltonian to the real space and introduce the repulsive Hubbard interaction term. In section \ref{sec-exact}, it is shown that the exact solutions of this model can be easily revealed by rewriting its Hamiltonian in terms of Majorana fermions \cite{shen2012topological,chen2018exactly}. We also provide detailed studies of the band structure and edge modes in both non-interacting limit and also the fullly interacting case. The Chern number and system energy of the interacting model are computed numerically.

\section{Constructing the topological superconductor with flat bands}
\label{sec-construct}

We will briefly review the Qi-Wu-Zhang model, which is prototype of two-dimensional Chern insulators on square lattice. In the momentum space, the Hamiltonian in second quantized form is given by
\be
\mathcal{H}_{CI}=\sum_{\vk}\sum_{a,b=1}^2\dc_{a,\vk} H_{ab}(\vk)c_{b,\vk},\quad H=\sum_{j=1}^3R_j\sigma_j
\ee
Here $a,b$ labels the two orbital in each site. $\sigma_j$ for $j=1,2,3$ are Pauli matrices. The three coefficients in front of the Pauli matrices is given by
\be
&&R_1=m+\cos k_x+\cos k_y,\quad R_2=\sin k_x,\quad R_3=\sin k_y
\ee
Since the hopping and potential terms of the above model do not depend on the spin, we can treat this model as a spinless fermion model. The Qi-Wu-Zhang model has no internal symmetries and belongs to the class A of the tenfold way classification of topological insulators \cite{Ryu2008:prb,Chiu2016:rmp}. In two-dimension, the topological property of class A is characterized by the $\mathbb{Z}$ classification, which is simply the Chern number of the occupied energy bands.
It is well known that for a two-band model, the Chern number can be written as a winding number of the mapping from $k$-space to a unit sphere \cite{Qi-TFT} as
\be
C=-\frac1{4\pi}\int d^2k\epsilon_{ijk}\hat{R}_i\cdot(\p_x\hat{R}_j\times\p_y\hat{R}_k)
\label{wind}
\ee
Here $R=\sqrt{R_1^2+R_2^2+R_3^2}$ and $\hat{R_i}=R_i/R$ is unit vector along $R_i$. Applying the above formula to the Qi-Wu-Zhang model, it is easy to see that the Chern number of lower band depends on the parameter $m$ as follows
\be
C=\left\{
    \begin{array}{ll}
      1, & 0<m<2\\
      -1, & -2<m<0 \\
      0, & |m|>2
    \end{array}
  \right.
\label{QWZ}
\ee
The non-trivial Chern number appears when all $R_i$ can change signs in the $k$-space.

In order to construct an exact solvable model from the above Chern insulator, we first extend the above model into a topological superconductor by adding certain pairing terms. The pairing terms can be treated as equal spin pairing, which have similar structure as the hopping terms. These pairing terms are specially constructed such that the model can support flat bands for some appropriate choice of parameters. In experiments, this type of pairing terms can be physical realized in the  superconductor-topological insulator heterostructure by  superconducting proximity effect \cite{FK2008:prl,Xu2014:prl,Sau2010:prb,Stanescu2010:prb}.
To express the superconductor model, we introduce the Nambu spinor as follows
\be
\psi_{\vk}=(c_{1,\vk},\,c_{2,\vk},\,\dc_{1,-\vk},\,\dc_{2,-\vk})^T
\ee
Then the Hamiltonian can be written as the standard Bogoliubov-de Gennes (BdG) form as
\be
\mathcal{H}_{TS}=\sum_{\vk}\sum_{ab}\psi^{\dag}_{a,\vk} \Big(H_{TS}(\vk)\Big)_{ab}\psi_{b,\vk},
\quad H_{TS}(\vk)=\left(
    \begin{array}{cc}
      \xi(\vk) & \Delta \phi(\vk) \\
      \Delta \phi^{\dag}(\vk) & -\xi(-\vk)^*
    \end{array}
  \right)
  \label{TS}
\ee
Here the kinetic term and pairing terms are given by
\be
\xi(\vk)=\sum_{j=1}^3R_j\sigma_j,\quad
\phi(\vk)=i (R_1\sigma_2- R_2\sigma_1)-R_3\sigma_0
\ee
where $\sigma_0$ is 2 by 2 identity matrix. The topological superconductor has the following particle-hole symmetry
\be
(\tau_1\otimes\sigma_0) H^*(\vk)(\tau_1\otimes\sigma_0)=-H(-\vk)
\ee
here $\tau_1$ is the Pauli matrix applying to the Numbu spinor space. Due to this particle-hole symmetry, this model belongs to the class D of tenfold way classification. In dimension two, the class D is also classified by integer group $\mathbb{Z}$, which is again labeled by the Chern number of the occupied energy bands.

The topological superconductor Hamiltonian in the first quantized form can be written more explicitly as a $4\times 4$ matrix
\be
H_{TS}=\left(
    \begin{array}{cccc}
      R_3 & R_- & -\Delta R_3 & \Delta R_- \\
      R_+ & -R_3 & -\Delta R_+ & -\Delta R_3 \\
      -\Delta R_3 & -\Delta R_- & R_3 & -R_- \\
      \Delta R_+ & -\Delta R_3 & -R_+ & -R_3
    \end{array}
  \right)
  \label{MTS}
\ee
Here $R_{\pm}=R_1\pm i R_2$. In this form, it is easy to compute the four energy bands, which are given by
\be
E=\pm(\Delta\pm1)\sqrt{R_1^2+R_2^2+R_3^2}
\ee
If one assumes that $\Delta=1$, there are two bands become exactly flat with $E=0$. The appearance of flat bands is crucial for construction of an exact solvable model with Hubbard interaction. Since the flat band fermions has zero energy, their fermion operators do not evolve with time and can be treated as constants. Therefore any four fermion interaction term involving two flat band fermions and two other fermions will reduce to a quadratic fermion term. This in turn makes the interacting model exact solvable. We will get back to this point in details in the next section.

Before introducing the Hubbard interaction term, we can first discuss the topological property of the above model. From now on, we will always assume that the flat band condition $\Delta=1$ is satisfied. This situation is also the focus of the next section. In this case, the eigenvector of the lowest band can be analytically expressed as
\be
\psi=\Big(-\frac{\sqrt{R(R-R_3)}}{2R},\,\frac{R_1+i R_2}{2\sqrt{R(R-R_3)}},
\,\frac{R-R_3}{2\sqrt{R(R-R_3)}},\,\frac{R_1+i R_2}{2\sqrt{R(R-R_3)}}\Big)^T
\ee
Plug the above result into the definition of Chern number
\be
C=\frac{-i}{2\pi}\int d^2k\Big(\ep{\p_x\psi|\p_y\psi}-\ep{\p_y\psi|\p_x\psi}\Big)
\ee
After some algebraic calculations, we find the same result of the Chern number as for a two-band model
\be
C=-\frac1{4\pi}\int d^2k\epsilon_{ijk}\hat{R}_i\cdot(\p_x\hat{R}_j\times\p_y\hat{R}_k)
\ee
Therefore, the Chern number of the model Eq.(\ref{TS}) is the same as the Chern number of the Qi-Wu-Zhang model in Eq.(\ref{QWZ}).

\begin{figure}
\centering
\includegraphics[width=\columnwidth]{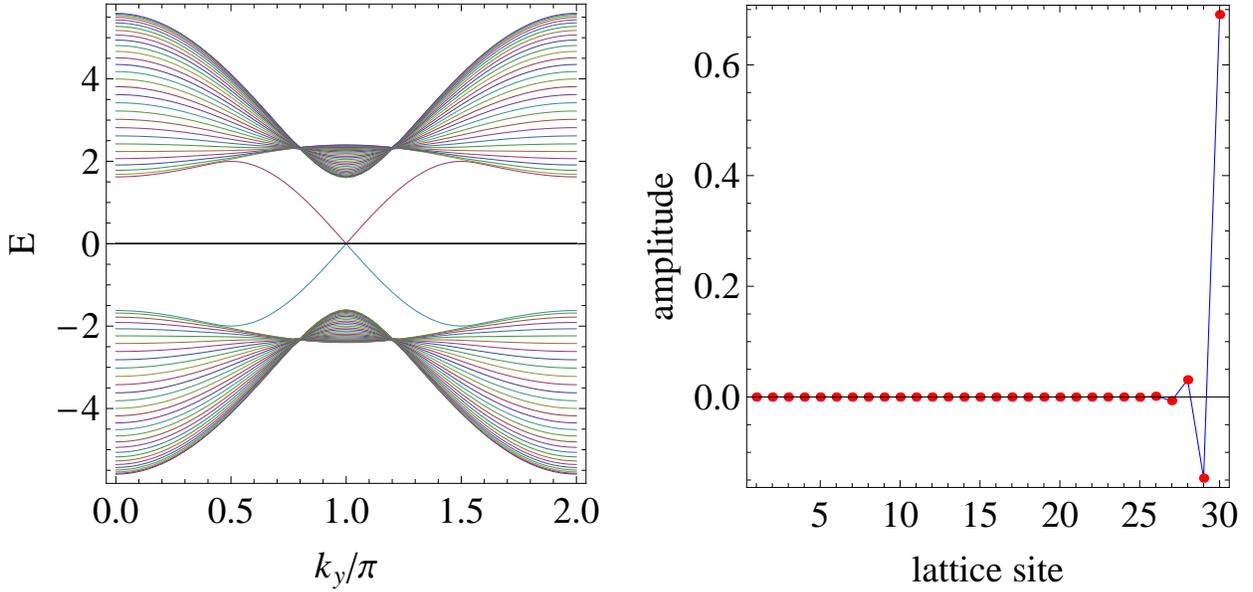}
\caption{Left panel: Energy eigenvalues of Eq.(\ref{TS}) as a function of $k_y$ with open boundary in $x$ direction with $m=0.8$ and $\Delta=1$. Right panel: The amplitude of an edge mode at $k_y=1.3\pi$ as a function of lattice sites.}
\label{fig-TS}
\end{figure}

The result of Chern number can also be reflected through the number of edge modes by the bulk-edge correspondence. To investigate the edge modes of the topological superconductor model of Eq.(\ref{TS}), we realize this model on a cylinder-shape lattice. To be concrete, we assume an open boundary condition along $x$ axes and keep the periodic boundary condition along $y$ axes. The resulting Hamiltonian is still a function of $k_y$. In the left panel of Figure \ref{fig-TS}, we plot the eigenvalues as a function of $k_y$. We have assumed that $\Delta=1$ which is the flat band condition, and $m=0.8$ corresponding to $C=1$. One can see that beside the two valence and conduction bands, there are two perfect flat bands located at zero energy. Other than the 4 bulk bands, there are two chiral edge modes connecting the valence and conduction bands. In the right panel of Figure \ref{fig-TS}, the amplitude of one of the edge modes is shown as a function of the lattice sites. One can see the amplitude is localized at one end of the cylinder, which confirms that this is indeed an edge mode.

As a side note, we mention that it is also convenient to rewrite the topological superconductor model in a more compact form as
\be
H=R_1\Gamma_{31}+R_2\Gamma_{32}+R_3\Gamma_{03}-\Delta(R_1\Gamma_{22}-R_2\Gamma_{21}+R_3\Gamma_{10})
\label{Ga}
\ee
Here we introduce $\Gamma_{ij}=\tau_i\otimes\sigma_j$ with $\tau_i$ acting on the Nambu spinor space and $\sigma_j$ acting on the orbital space.

\section{Introducing the Hubbard interaction}
\label{sec-Hubbard}

In order to introduce the four-fermion Hubbard interaction term, we extend the topological superconductor model Eq.(\ref{TS}) to include the spin degree of freedom. In the first quantized form, the Hamiltonian can be written as
\be
H=H_{TS}\otimes s_0
\ee
Here $s_0$ is a $2\times 2$ identity matrix acting on the spinor space and $H_{TS}$ is from Eq.(\ref{MTS}).

Since the Hubbard interaction is an on-site repulsion in real space, we can also transfer the hopping and pairing terms of the above model to the real space as follows
\be
&&H_{\textrm{CI}}=2\sum_{n,s}\Big(\dc_{n,s}\frac{\sigma_1-i\sigma_2}{2}c_{n+\hat{x},s}
+\dc_{n,s}\frac{\sigma_1-i\sigma_3}{2}c_{n+\hat{y},s}+h.c.\Big)
+2\sum_{n,s} m \dc_{n,s} \sigma_1 c_{n,s}\label{CI}\\
&&H_{\textrm{pair}}=2\Delta\sum_{n,s}\Big(\dc_{n,s}\frac{i\sigma_2-\sigma_1}2\dc_{n+\hat{x},s}
+\dc_{n,s}\frac{i\sigma_2+i\sigma_0}2\dc_{n+\hat{y},s}+h.c.\Big)\nonumber\\
&&\qquad+\Delta\sum_{n,s}\Big(m \dc_{n,s}(i\sigma_2) \dc_{n,s}+h.c.\Big)\label{pair}
\ee
Here $n=(n_x,n_y)$ labels the lattice site on a square lattice and $s=1,2$ labels the spin up and down. The two-component fermion operator is defined as $c_{n,s}=(c_{n,1,s}, c_{n,2,s})^T$. The second index labels the two orbital. We also introduce $\hat{x}$ and $\hat{y}$ representing the unit vector along the $x$ and $y$ direction respectively. The spin dependence of above Hamiltonian is trivial, because the spin up and down behave exactly the same. Nevertheless, we still keep the spin index explicit in the above equations, because the Hubbard interaction term we are about to introduce will depend on the spin.

To verify the correctness of Eq.(\ref{CI}) and Eq.(\ref{pair}), we can make the following fourier transformation back to momentum space,
\be
c_{n,j,s}=\frac{1}{\sqrt{L_x L_y}}\sum_{\vk}c_{\vk,j,s}\exp(i\,\vk\cdot\vn)
\ee
Here $L_x$ and $L_y$ are the lattice site number along the $x$ and $y$ axes. For the hopping terms, it is straightforward to see that the transformation gives back the Qi-Wu-Zhang model in momentum space. For the pairing terms, we make use of the following identities
\be
&&\sum_{n,ij}\dc_{n,i}M_{ij}\dc_{n+\hat{x},j}=\sum_{\vk,ij}\dc_{\vk,i}M_{ij}\dc_{-\vk,j}\cos k_x,
\quad\mbox{for}\quad M^T=-M\\
&&\frac1{i}\sum_{n,ij}\dc_{n,i}M_{ij}\dc_{n+\hat{x},j}=\sum_{\vk,ij}\dc_{\vk,i}M_{ij}\dc_{-\vk,j}\sin k_x,
\quad\mbox{for}\quad M^T=M
\ee
Then it is easy to see that Eq.(\ref{pair}) reproduce the desired pairing terms in momentum space.

Now we are in the position to define the Hubbard interaction term. It is an on-site repulsive interaction between spin up and spin down fermions. Explicitly, it can be written as
\be
H_{\textrm{int}}=U\sum_{n,j}\Big(\dc_{n,j,\uparrow} c_{n,j,\uparrow}-\frac12\Big)
\Big(\dc_{n,j,\downarrow} c_{n,j,\downarrow}-\frac12\Big)
\ee
For $U>0$, the zero and double occupation of each orbital will be punished by raising the system energy with $U/4$. On the other hand, the single occupation is favored by lowering the system energy with $U/4$.

Collect all the above results, we have arrived at the full interacting topological superconductor Hamiltonian in real space
\be
H=H_{\textrm{CI}}+H_{\textrm{pair}}+H_{\textrm{int}}
\label{full}
\ee
In the next section, we will show that this model can be exactly solved when the flat band condition $\Delta=1$ is satisfied.

\section{Exact solutions of the interacting model}
\label{sec-exact}

The exact solvability of the interacting model Eq.(\ref{full}) can be best understood by introducing the Majorana fermion operators as follows
\be
&&c_{n,1,s}=a_{n,1,s}+i b_{n,1,s},\quad \dc_{n,1,s}=a_{n,1,s}-i b_{n,1,s}\\
&&c_{n,2,s}=b_{n,2,s}+i a_{n,2,s},\quad \dc_{n,2,s}=b_{n,2,s}-i a_{n,2,s}
\ee
Here $s$ labels the spin up and down. It is easy to see that they satisfy the following anti-commutation relations
\be
\{a_{m,j,s},a_{n,j',s'}\}=\{b_{m,j,s},b_{n,j',s'}\}=\frac12\delta_{mn}\delta_{jj'}\delta_{ss'}
\ee
Here $m,n$ labels the lattice sites, $j,j'$ labels the two orbital, $s,s'$ labels the spin degree of freedom.

Substituting the Majorana representation into Eq.(\ref{full}), after some algebraic calculation, we find that the hopping and pairing terms can be written as
\be
&&H_{\textrm{CI}}+H_{\textrm{pair}}=2i\sum_n\Big[2(\Delta_1b_{n,2}b_{n+\hat{x},1}-\Delta_2 a_{n,2}a_{n+\hat{x},1})
+(\Delta_2 a_{n,1}a_{n+\hat{y},2}-\Delta_1 b_{n,1}b_{n+\hat{y},2})\nonumber\\
&&\quad+(\Delta_1 b_{n,2}b_{n+\hat{y},1}-\Delta_2 a_{n,2}a_{n+\hat{y},1})
-(\Delta_2a_{n,1}a_{n+\hat{y},1}+\Delta_1 b_{n,1}b_{n+\hat{y},1})\nonumber\\
&&\quad+(\Delta_1 b_{n,2}b_{n+\hat{y},2}+\Delta_2 a_{n,2}a_{n+\hat{y},2})
+2(\Delta_2 a_{n,1}a_{n,2}-\Delta_1 b_{n,1}b_{n,2})\Big]
\label{Majorana}
\ee
Here $\Delta_1=1+\Delta$ and $\Delta_2=1-\Delta$. In the meanwhile, he Hubbard interaction can be expressed as
\be
H_{\textrm{int}}=U\sum_{n,j}(2ia_{n,j,\uparrow}b_{n,j,\uparrow})(2ia_{n,j,\downarrow}b_{n,j,\downarrow})
\ee
When the flat band condition $\Delta=1$ is satisfied, all kinetic terms for $a_{n,j,s}$ in Eq.(\ref{Majorana}) are vanished. If we define $D_{n,j}=4ia_{n,j,\uparrow}a_{n,j,\downarrow}$, then it is easy to see the $[D_{n,j},H]=0$ for all sites $n$ and $j=1,2$, which makes $D_{n,j}$ a c-number. From the anti-commutation relations, we find $D_{n,j}^2=1$, therefore $D_{n,j}=\pm 1$. Suppose the total number of lattice sites is $N$, then there are $2^{2N}$ choices of different configurations of $D_{n,j}$. For a given configuration of $D_{n,j}$, the Hubbard interaction term reduces to a quadratic term of Majorana fermions $b_{n,j,s}$. Therefore, the full interacting model becomes quadratic in fermion operators, which can be easily solved.

\begin{figure}
\centering
\includegraphics[width=\columnwidth]{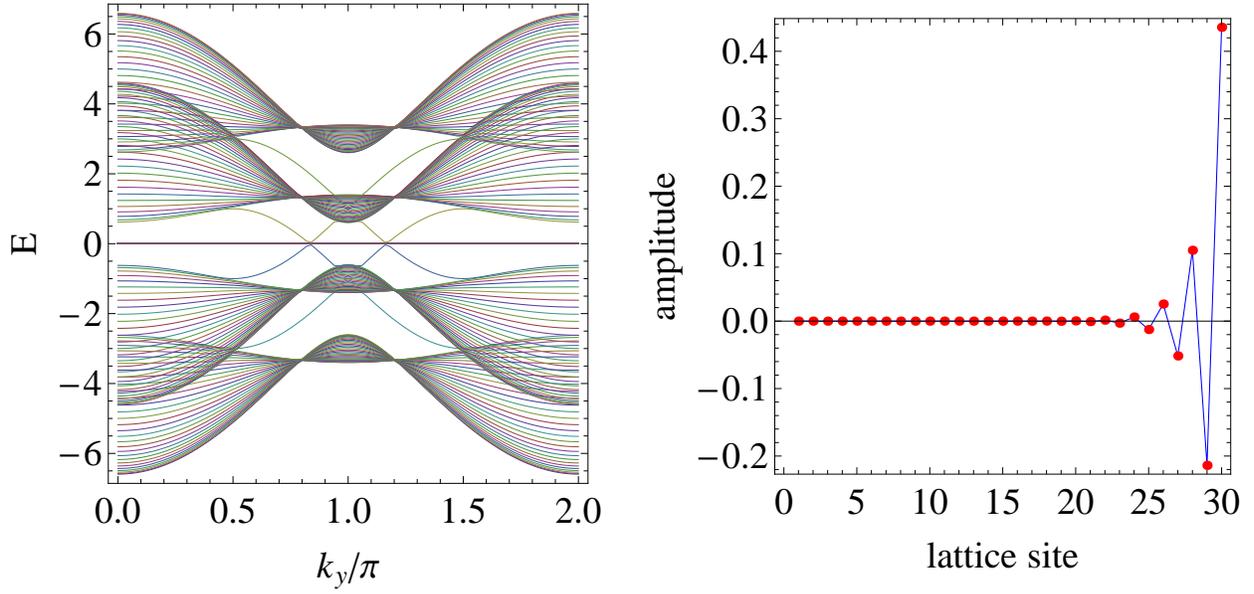}
\caption{Left panel: Energy eigenvalues as a function of $k_y$ with open boundary in $x$ direction for $U=2$, $m=0.8$ and $\Delta=1$ with uniform distribution of $D_n$. Right panel: The amplitude of an edge mode at $k_y=1.4\pi$ as a function of lattice sites.}
\label{fig-U1}
\end{figure}

\begin{figure}
\centering
\includegraphics[width=\columnwidth]{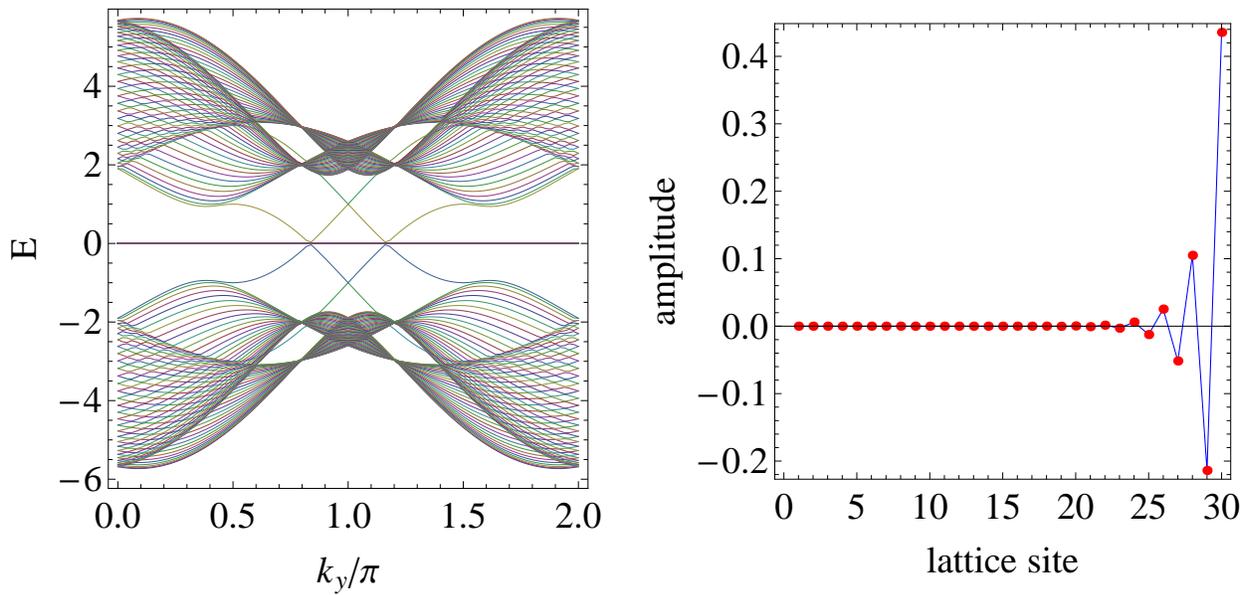}
\caption{Left panel: Energy eigenvalues as a function of $k_y$ with open boundary in $x$ direction for $U=2$, $m=0.8$ and $\Delta=1$ with staggered distribution of $D_n$. Right panel: The amplitude of an edge mode at $k_y=1.4\pi$ as a function of lattice sites.}
\label{fig-U2}
\end{figure}

Among all the configurations of $D_{n,j}$, there are two regular configurations obviously require more detailed examination. In contrast to the random configuration which can only be solved in real space, the regular configuration allows one to transfer the Hamiltonian back to momentum space, which greatly simplifies the computation of eigenvalues. One of the regular configurations is the uniform distribution with $D_{n,j}=1$ for all sites and orbital. If we transfer to the momentum space and rewrite the Majorana fermion back to ordinary fermion operators, we find that the Hubbard interaction term can be written as
\be
H_U=-\frac{U}{4}(\Gamma_{00}\otimes s_2-\Gamma_{13}\otimes s_2)
\ee
Here $\Gamma_{ij}=\tau_i\otimes\sigma_j$ with $\tau_i$ acting on the Nambu spinor space and $\sigma_j$ acting on the orbital space. The matrices $s_j$ acts on the spin space. In this case, the 4 energy bands can be simple expressed as
\be
E_U=\pm2\sqrt{R_1^2+R_2^2+R_3^2}\pm\frac{U}{2}
\ee
The other 4 are flat bands with $E=0$.

The other regular configuration is the staggered distribution with $D_{n,1}=1$ and $D_{n,2}=-1$ for all sites. Transferring back to the original fermion operators, the Hubbard interaction can be written as
\be
H_S=-\frac{U}{4}(\Gamma_{03}\otimes s_2-\Gamma_{10}\otimes s_2)
\ee
In this case, 4 out of the 8 energy bands can be written as
\be
E_S=\pm2\sqrt{R_1^2+R_2^2+(R_3\pm U/4)^2}
\ee
and the other 4 are zero energy flat bands. We will see later that the staggered configuration is the true ground state of the full interacting model.

Now we can investigate the band structure and edge modes of the interacting model. To this end, we can again realize the above model on a cylinder geometry with open boundary along $x$ axes. Then one can numerically diagonalize the resulting Hamiltonian which depends on the remaining momentum $k_y$. For the uniform configuration, we plot the eigenvalues as a function of $k_y$ in the left panel of Figure \ref{fig-U1}. One can see that the interaction term has lifted the degeneracy between the spin up and down fermions. Thus, there are two valence bands below $E=0$ and two conduction bands above zero energy. Each band can be mapped to itself by reflecting about $k_y=\pi$. There are clearly two pairs of edge modes signaling the non-trivial topology of these bulk bands. The wave-function of a typical edge mode is shown in the right panel of Figure \ref{fig-U1}. Its amplitude is localized at the right end, confirming it is truly an edge mode. Similarly, we show the eigenvalues as a function of $k_y$ for the staggered configuration in the left panel of Figure \ref{fig-U2}. Again, there are 4 different energy bands for the two orbital and two spins. In contract to the uniform case, for the staggered case, the two valence bands can be mapped to each other by reflecting about $k_y=\pi$. There also appears two pairs of chiral edge modes with its wave-function shown in the right panel of Figure \ref{fig-U2}.

The number of edge modes is also reflected by the Chern number of the bulk bands. For the interacting model, the eigenvectors are too complicated to support an analytical expression, thus it is impossible to find a simple result of Chern number as in Eq.(\ref{wind}). But it is still straightforward to compute the Chern number numerically for each separate bands. For the fixed interaction strength $U=2$, the numerical result shows that for the uniform case, the Chern numbers of the lower two bands are
\be
C_1=C_2=\textrm{sign}(m)\Theta(2-|m|)
\ee
Here the $C_i$ with $i=1,2$ labels the lowest and second lowest band. $\Theta(x)$ is the step function. For the staggered case, the Chern numbers are
\be
C_1=0,\quad C_2=2\textrm{sign}(m)\Theta(2-|m|)
\ee
The above results of Chern number indeed match the number of edge modes discussed above.

\begin{figure}
\centering
\includegraphics[width=0.5\columnwidth]{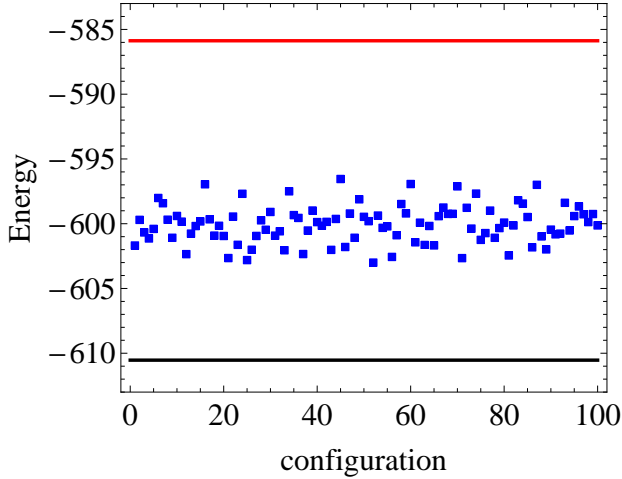}
\caption{The blue squares represent the  system energies for 100 disordered configurations. The black thick line represents the energy of the staggered configuration. The red thick line represents the energy of the uniform configuration.}
\label{fig-gs}
\end{figure}

At last, we want to study the energetics of the model in order to determine which configuration is the ground state of the system. The model is realized on a $10\times 10$ lattice with open boundaries along both directions. Then we randomly choose $D_{n,j}$ and numerically find out all eigenvalues. The system energy is obtained by the summation of all negative eigenvalues. For such a large system, it quite expensive to enumerate all possible disordered configurations. Therefore, in Figure \ref{fig-gs}, we plot the system energies for 100 randomly chosen disordered configurations. The black and red thick lines corresponds the energies of the staggered and uniform configuration. One clear see that staggered configuration is the true ground state of the system. On the other hand, the uniform configuration gives rise to the highest system energy.

\section{Conclusion}

We have constructed a topological superconductor with equal spin pairing and Hubbard repulsive interaction, which can be exactly solved when the pairing gap equals to the hopping constant. The topology of this interacting model is examined in detail. The edge modes are displayed explicitly by putting the model on a cylinder geometry and numerically solve the band structure. It is found that the number of edge modes exactly matches the Chern number of occupied bands, which is consistent with the bulk-edge correspondence. Finally, the true ground state with staggered configuration is determined by comparison with many random configurations.

\begin{acknowledgments}
This work is supported by the National Natural Science Foundation of China under Grant No. 11874272 and Science Specialty Program of Sichuan University under Grant No. 2020SCUNL210.
\end{acknowledgments}


%

\end{document}